\documentclass[amsmath,amssymb,superscriptaddress,nobalancelastpage,aps,prl,twocolumn]{revtex4-2}

\usepackage{graphicx}
\usepackage{varioref}
\usepackage{xr-hyper}
\usepackage{xcolor}
\usepackage{nicefrac}
\usepackage{xfrac}
\usepackage{hyperref}
\hypersetup{colorlinks,linkcolor=blue,urlcolor=blue,citecolor=blue}
\usepackage{ulem}
\usepackage{lineno}
\usepackage{amsmath} 
\usepackage{amssymb}
\usepackage{siunitx}

\newcommand{\Ca}{Ca$_{2}$RuO$_{4}$}

\begin{document}

\title{Resolving the Orbital Character of Low-energy Excitations in Mott Insulator with Intermediate Spin-orbit Coupling}
\author{K.~von~Arx}
\affiliation{Department of Physics, Chalmers University of Technology, SE-412 96 G\"{o}teborg, Sweden}
\affiliation{Physik-Institut, Universit\"{a}t Z\"{u}rich, Winterthurerstrasse 
190, CH-8057 Z\"{u}rich, Switzerland}

\author{P. Rothenbühler}
\affiliation{Physik-Institut, Universit\"{a}t Z\"{u}rich, Winterthurerstrasse 190, CH-8057 Z\"{u}rich, Switzerland}

\author{Qisi~Wang}
\affiliation{Physik-Institut, Universit\"{a}t Z\"{u}rich, Winterthurerstrasse 
190, CH-8057 Z\"{u}rich, Switzerland}
\affiliation{Department of Physics, The Chinese University of Hong Kong, Shatin, Hong Kong, China}

\author{J.~Choi}
\affiliation{Diamond Light Source, Harwell Campus, Didcot, Oxfordshire OX11 0DE, UK}

\author{M.~Garcia-Fernandez}
\affiliation{Diamond Light Source, Harwell Campus, Didcot, Oxfordshire OX11 0DE, UK}

\author{S.~Agrestini}
\affiliation{Diamond Light Source, Harwell Campus, Didcot, Oxfordshire OX11 0DE, UK}

\author{Ke-Jin~Zhou}
\affiliation{Diamond Light Source, Harwell Campus, Didcot, Oxfordshire OX11 0DE, UK}

\author{A.~Vecchione}
\affiliation{CNR-SPIN, I-84084 Fisciano, Salerno, Italy}
\affiliation{Dipartimento di Fisica ``E.R.~Caianiello", Universit\`{a} di Salerno, I-84084 Fisciano, Salerno, Italy}

\author{R.~Fittipaldi}
\affiliation{CNR-SPIN, I-84084 Fisciano, Salerno, Italy}
\affiliation{Dipartimento di Fisica ``E.R.~Caianiello", Universit\`{a} di Salerno, I-84084 Fisciano, Salerno, Italy}

\author{Y.~Sassa}
\affiliation{Department of Physics, Chalmers University of Technology, SE-412 96 G\"{o}teborg, Sweden}

\author{M.~Cuoco}
\affiliation{CNR-SPIN, I-84084 Fisciano, Salerno, Italy}
\affiliation{Dipartimento di Fisica ``E.R.~Caianiello", Universit\`{a} di Salerno, I-84084 Fisciano, Salerno, Italy}

\author{F.~Forte}
\email{filomena.forte@spin.cnr.it}
\affiliation{CNR-SPIN, I-84084 Fisciano, Salerno, Italy}
\affiliation{Dipartimento di Fisica ``E.R.~Caianiello", Universit\`{a} di Salerno, I-84084 Fisciano, Salerno, Italy}

\author{J.~Chang}
\affiliation{Physik-Institut, Universit\"{a}t Z\"{u}rich, Winterthurerstrasse 190, CH-8057 Z\"{u}rich, Switzerland}
\maketitle
  \textbf{Multi-band Mott insulators with moderate spin-orbit and Hund’s coupling are key reference points for theoretical concept developments of correlated electron systems. The ruthenate Mott insulator \Ca\ has therefore been intensively studied by spectroscopic probes. However, it has been challenging to resolve the fundamental excitations emerging from the hierarchy of electronic energy scales. Here we apply state-of-the-art resonant inelastic x-ray scattering to probe deeper into the electronic excitations found in \Ca. In this fashion, we probe a series of spin-orbital excitations at low energies and resolve the level splitting of the intra-$t_{2g}$ structure due to spin-orbit coupling and crystal field splitting. Most importantly, the low-energy excitations exhibit strong orbital character. Such direct determination of relevant electronic energy scales is important, as it sharpens the target for theory developments of Mott insulators’ orbital degree of freedom.}
  

\section{Introduction}
Materials with moderate to strong spin-orbit coupling (SOC) have long been the center of intense research~\cite{WitczakKrempa2014}. In some material classes – such as the iridates – SOC is a key driver of multi band Mott physics~\cite{KimPRL2008}. In metals, SOC is an important factor for problems linked to topology~\cite{Pesin2010,Bernevig2006}. A third pillar is the combination of SOC and magnetism~\cite{ChenPRB2008,KunkemollerPRL2015}. In such systems, spin and orbital excitations can no longer be treated separately. This setting offers a ground for new types of collective excitations but also adds further layers of complexity to both theory and experimental efforts.
On the experimental side, resonant inelastic x-ray scattering (RIXS) offers a unique possibility by being sensitive to both spin and orbital excitations~\cite{amentRMP2011}. However, only few systems with magnetic excitations also have a sizeable SOC. Most often, these energy scales are inaccessible to RIXS due to modest energy resolution. Recently, advances in soft RIXS methodology has greatly improved the energy resolution~\cite{Zhou2022}. This yields new opportunities to tackle the problem of spin-orbit coupled magnetic ground states. 



For example, in the archetypal multi-band Mott insulator \Ca~\cite{GorelovPRL2010,han_lattice_2018,Ricc2018,ZhangPRX2019,NakatsujiPRL2004}, many questions related to low-energy scales are still open. Previous RIXS studies~\cite{das_spin-orbital_2018,GretarssonPRB2019,FatuzzoPRB2015} have revealed a series of excitations emerging from the interplay of Coulomb interaction $U$, Hund’s coupling $J_H$, crystal-field splitting $\delta$ and SOC $\lambda$. It has been shown that SOC is a crucial element to explain the nature of these excitations and the activation in the RIXS process. However, theoretical modelling of the RIXS spectra only allows for an estimation of the relative strength of crystal-field and SOC, but the value of SOC is imprinted in the fine structure of an excitation sector. This fine structure, as well as a spin excitation predicted by the model around 40 meV, could previously not be resolved due to limited energy resolution.


\begin{figure*}[t]
 	\begin{center}
 		\includegraphics[width=0.99\textwidth]{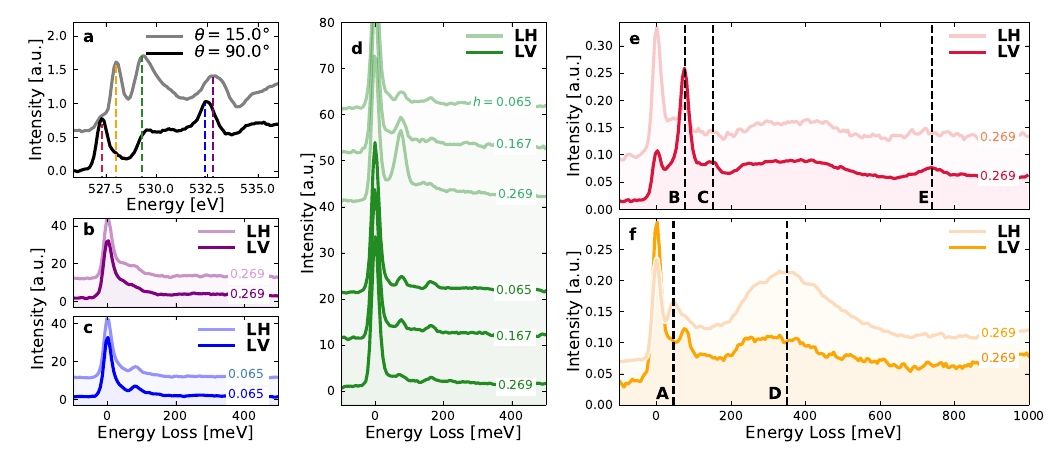}
 	\end{center}
 	\caption{ \textbf{X-ray absorption spectroscopy (XAS) and resonant inelastic x-ray scattering (RIXS) on \Ca.} (a) X-ray absorption spectra at the oxygen $K$-edge recorded with LH polarized light and incidence angle as indicated. Dashed colored lines indicate the absorption resonances used to collect RIXS spectra in (b-f).
 	(b-d) RIXS spectra (intensity versus energy loss) obtained on the $e_g$ resonances for light polarizations and momentum $\boldsymbol{q}=(h,0)$ as indicated. (e,f) Spectra recorded on the $t_{2g}$ resonances.  Vertical dashed lines in (e) and (f) indicate the observed excitations -- labeled A, B, C, D, and E.  The color code in (b-f) matches the resonance identification (vertical dashed lines) in panel (a).}
	 \label{fig:fig1}
\end{figure*}

Here we present state-of-the-art high-resolution oxygen $K$-edge RIXS measurements on \Ca, enabling a deeper insight into the low-energy structure of this SOC-driven multiband Mott insulator. This approach allows for new findings. In particular, the 40 meV magnetic mode was observed for the first time by RIXS, and a dispersive nature of the 80 meV mode is revealed. Our theoretical modelling reveals a complex spin-orbital origin of these modes, having a dominant transverse spin component with distinct orbital character, making them accessible at the O $K$-edge by modest SOC. Most importantly, the orbital character of excitation modes corresponding to longitudinal and transverse variation of the orbital angular momentum is inferred. 
Additionally, excitations at 160, 350, and 750 meV are observed. The fine structure (energy level splitting) of the 350 meV excitation revealed direct insight into the SOC. 
Thus, the improvement in energy resolution allows for a detailed observation of the excitation spectrum and the precise extraction of the SOC fine structure, which is essential for understanding the complex and rich properties of \Ca.

\begin{figure*}[t]
 	\begin{center}
 	\includegraphics[width=0.99\textwidth]{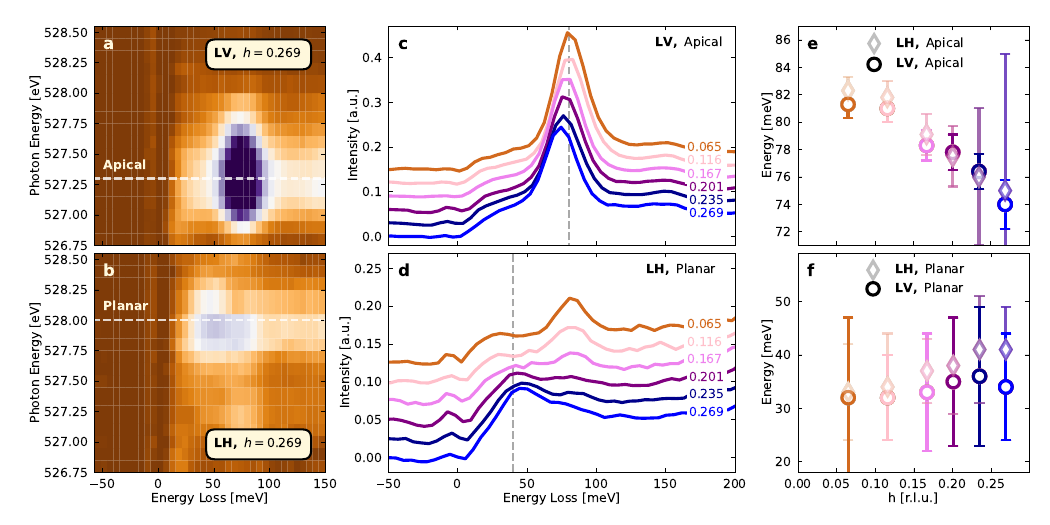}
 	\end{center}
 	\caption{\textbf{Low-energy RIXS spectra recorded on \Ca.} (a,b) RIXS spectra as a function of incident photon-energy (vertical axis) for linear vertical and horizontal light and momentum as indicated. Horizontal dashed lines mark respectively the planar and apical oxygen pre-edges. Intensity is displayed using a false color scale. (c,d) RIXS spectra versus momentum as indicated. Vertical dashed lines are guides to the eye for the A and B excitations.
 	To highlight the low-energy excitations,  elastic scattering is subtracted in (a)-(d). The dispersions -- obtained from fits shown in supplementary Fig. xx -- of the A and B excitations are shown in panel e,f. From (c),(e), a significant dispersion of the B excitation is resolved. Within error bars no dispersion is derived for the A excitation. }
	 \label{fig:fig2}
\end{figure*}

\begin{figure}[h!]
 	\begin{center}
 		\includegraphics[width=0.48\textwidth]{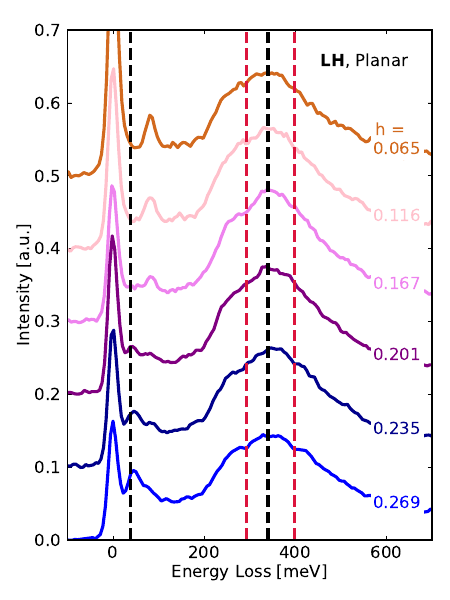}
 	\end{center}
 	\caption{ \textbf{Resolved energy level splitting due to spin-orbit coupling.} Spectra focusing on the excitation around 350 meV, recorded at different in-plane momenta as indicated. Black vertical dashed lines represent the excitation values extracted from the data used to calculate crystal field and SOC. Red vertical dashed lines represent the excitation values obtained with the theoretical model. }
	 \label{fig:fig3}
\end{figure}

\noindent\textbf{Results}\\
Fig.~\ref{fig:fig1}a displays oxygen $K$-egde x-ray absorption spectra recorded with linear horizontal polarized light for different angles $\theta$ between sample surface and incident light as indicated. The two pre-edges originating from the hybridisation of the oxygen orbitals with the ruthenium  $t_{2g}$ states split due to the different crystal field environment around the planar and apical oxygen sites. These $t_{2g}$ resonances are followed by broader $e_g$ resonances. RIXS spectra are collected at both the $t_{2g}$ and $e_{g}$ resonances - see Fig.~\ref{fig:fig1}b-f using both linear vertical (LV) and horizontal (LH) light polarizations.
In total, five (low-energy) excitations, labeled A, B, C, D and E, are found in the $t_{2g}$ spectra. The $e_g$ and $t_{2g}$ spectra both display excitations at 80 meV and 160 meV (B,C). We stress that excitations B, D, and E at 80,~350, and 750 meV, have been reported in previous RIXS studies~\cite{FatuzzoPRB2015,das_spin-orbital_2018,GretarssonPRB2019}. Yet, with increased energy resolution we reveal new information about these excitations. Excitation A and C are to the best of our knowlegde reported here for the first time. The energy scale of the A excitation matches with the magnetic amplitude mode reported by neutron scattering~\cite{jain2017higgs}. 
In this context, it is worth noticing that excitation B is found at twice the energy of excitation A, excitation C at twice the energy of B and these three excitations are observed at the same absorption resonances and light polarizations. 

The A and B excitations - after subtracting the elastic scattering profile - are shown in Fig.~\ref{fig:fig2}. The B excitation is most pronounced in the LV channel at the apical oxygen site where it manifests strongly for all $h$ measured along $(h,0)$ (see Fig.~\ref{fig:fig2}a,c). The A excitation is most clearly observed with LH polarization at the planar oxygen site (see Fig.~\ref{fig:fig2}b,d) for large momentum transfers. Interestingly, the intensity amplitude of this excitation decays as the zone center is approached. By contrast, the B excitation shows the opposite trend under these experimental conditions, being strongest at the zone center.
Momentum dependence analysis of the A and B excitations reveal that B is dispersing clearly to lower energies upon increasing momentum.
With our applied energy resolution, we cannot resolve any significant dispersion of the A excitation.

Next, the results of the D excitation are discussed. The previous experimental data has been interpreted as a block of two excitations~\cite{das_spin-orbital_2018}. With the improved energy resolution, it is now clear that this block has at least three components, as can be seen for different momenta in Fig.~\ref{fig:fig3}. The theoretical model introduced earlier~\cite{das_spin-orbital_2018} predicts a splitting of the energy level around 350 meV in 4 components due to spin-orbit coupling. The assignment of the RIXS features is well understood in the simplified atomic picture, where all the low-energy states are written as linear combinations of the $|L_z,S_z\rangle$ vectors, expressed in terms of the $z$ projections of the $S=1$ and $L=1$ moments within the $ t_{2g}^4$ Ru manifold. In this framework, the lowest excitation contributing to the 350 meV structure is ascribed to the atomic transition between the ground state and the excited state $|$D$_1\rangle=\alpha|1,-1\rangle-\beta |-1,1\rangle$, with specific expression of the $\alpha$ and $\beta$ coefficient depending on the tetragonal crystal field splitting $\Delta$ and the spin-orbit coupling $\lambda$~\cite{das_spin-orbital_2018}.\\
For O $K$-edge RIXS, the local scattering process is spin-conserving because of the absence of spin-orbit coupling in the core hole intermediate state.
It turns out that the transitions induced at the Ru site between the dominant part of the spin-orbital configuration that contributes to the magnetic ground state (being mostly due to the $|0,0\rangle$ state) and the D$_1$ state are not allowed, since they correspond to $\Delta S_z=\pm 1$~\cite{das_spin-orbital_2018}.
The only non-vanishing dipole allowed atomic transitions to the $D_1$ state come from initial configurations $|1,-1\rangle$ and $|-1,1\rangle$, which are spin-orbit activated in the atomic ground state. However, those transitions have modest amplitudes and hence are not seen in the RIXS spectrum.

Thus, the three features seen in the RIXS spectrum can be associated to the higher energy levels D$_{2-4}$.
The data allows for a precise extraction of the energies for excitation A (Fig.~\ref{fig:fig2}f) and the middle level D$_3$, see black dashed lines in Fig.~\ref{fig:fig3}. With the theoretical model, these two energy levels allow for a calculation of the crystal field $\delta = 250$ meV and spin-orbit coupling $\lambda = 85$ meV. Furthermore, these extracted energy scales can now be inserted back into the model, allowing for a self-consistency check by calculating the excitation energy for the other two D levels observable by RIXS, illustrated by the red dashed lines. The predicted values agree well with the observed peak positions. Thus, the high-resolution data presented here allow for a more direct extraction of the spin-orbit coupling than previous studies.

\begin{figure}[h!]
 	\begin{center}
 	\includegraphics[width=0.46\textwidth]{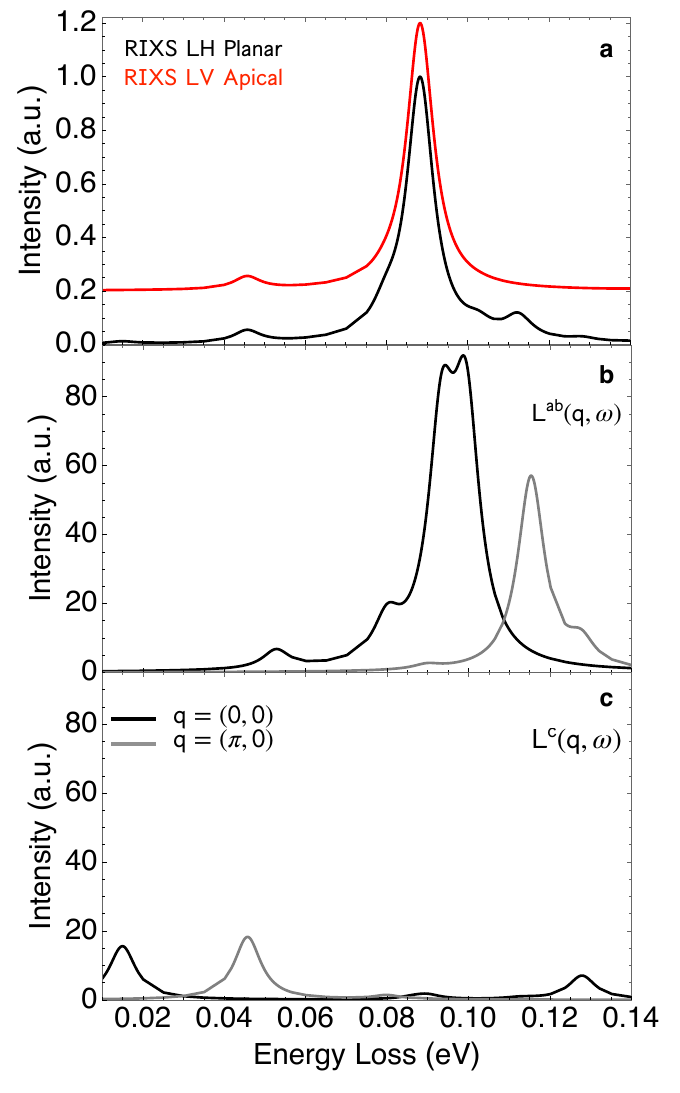}
 	\end{center}
 	\caption{ \textbf{Theoretical RIXS and dynamical orbital structure factor spectra}(a) Calculated RIXS spectra at $q=(0,0)$, for LH planar and LV apical resonance, for the finite size cluster as described in the text. The spectra were broadened with a 4 meV full width at half maximum (FWHM) Gaussian. (b) Dynamical orbital angular momentum structure factor $L^{ab}(q, \omega)$, related to the component of the orbital angular momentum in the $(a,b)$ plane (plane of the magnetic easy axis). (c) The dynamical orbital structure factor $L^{c}(q, \omega)$ related to the component of the orbital angular momentum along $c$ direction.}
	 \label{fig:fig4}
\end{figure}

\noindent\textbf{Discussion}\\
Starting from the low energy side of the excitation spectrum, this work presents the detection of the 40 meV excitation mode by RIXS. Previously, an inelastic neutron scattering study reported a dispersive Higgs mode with a similar energy scale at the zone center~\cite{jain2017higgs}. Additionally, a theoretical framework from recent RIXS studies~\cite{das_spin-orbital_2018, von_arx_resonant_2020} predicted a magnetic mode at approximately 40 meV, which should be observable through the RIXS cross-section. However, this energy range was previously not accessible in experiments due to limited instrumental energy resolution. This meant that no information about the orbital character of this excitation could be obtained, since neutron scattering is not sensitive to this orbital degree of freedom. With the state-of-the-art high energy resolution applied in this study, it is now possible to confirm and resolve this mode by RIXS.

Before discussing this in greater detail, we stress that the 40, 80 and 160~meV excitations appear with doubling energy scales - potentially suggesting a connected origin. A phononic nature of especially the 80 meV excitations should however be considered.
The 80 meV excitation appears very strongly in the LV channel on the apical resonance (Fig.~\ref{fig:fig1}e). In many oxide systems (for example cuprates~\cite{LinPRL2020,WangSCIADV2021} and titanates~\cite{GeondzhianPRL2020})
optical phonon branches have been observed in an energy range of 60 - 90 meV. 
Consistent with this expectation, Raman spectroscopy observe the high energy phonon in \Ca\ just below 80 meV~\cite{souliou_raman_2017}. From the extracted dispersion (Fig.~\ref{fig:fig2}e), the excitation at the zone center exceeds 80 meV. This is a first indication that this excitation is not consistent with a phononic nature. Its dispersion is also larger than expected for an optical phonon - a second indication. Finally, observation of a phonon in the RIXS spectra requires a strong electron phonon coupling which in turn generates intense higher-order harmonic excitations. The intensity ratio of the 80 and 160 meV excitations (Fig.~\ref{fig:fig2}c) does not suggest a strong electron phonon coupling~\cite{ValePRB2019,Dashwood2021}. This observation is the third evidence speaking against the phonon scenario. Based on these plausibility arguments, we interpret the RIXS spectra through an electronic model.   


Let us now consider the spin and orbital character of the low energy excitations (below 100 meV) and their relation to the different polarization channels at the planar and apical oxygen sites. 
The first question that we face is whether such low energy excitations are detectable by RIXS at the $K$-edge at both the apical and planar oxygen resonances.
In order to do that, we consider a microscopic model that includes the Coulomb interactions and the SOC at the Ru sites, as well as the symmetry allowed hybridization processes between Ru and O orbitals, in a regime for the microscopic parameters that is relevant for the Ca$_2$RuO$_4$~\cite{von_arx_resonant_2020}. 
The model Hamiltonian is then adapted to a finite size cluster consisting of two apical oxygens ($O_{\text{a}}$) and one central planar ($O_{\text{p}}$) oxygen site, thus having the following structure $O_{\text{a}}-Ru-O_{\text{p}}-Ru-O_{\text{a}}$. This cluster structure allows us to access the RIXS cross section at the planar and apical oxygen within the same simulation.
The determination of the RIXS cross section at the planar and apical oxygen atoms shows that the features at about 40 meV and 80 meV can be observed both at the apical and planar sites (see Fig.~\ref{fig:fig4} a). These outcome can account for the fact that in LV apical and LH planar spectra (Fig. 2) one can observe non-vanishing peaks around 40 and 80 meV.
\\
Concerning the character of these excitations, it is useful to consider the dynamical spin and orbital response  described by the structure factors $S(q,\omega)$ and $L(q,\omega)$, related to the localized spin and orbital angular momenta  of the $d$-electrons at the Ru site.
For our purposes, the analysis of the orbital dynamics is particularly relevant because the weight of the RIXS spectra exhibits a sizable dependence on the light polarization, incidence angle and on the chosen (planar or apical) oxygen resonance. 
First, we recall that the magnetic ground-state has anisotropic antiferromagnetic correlations among the Ru spins with moments mostly lying within the $ab$ plane ~\cite{porterPRB2018,fortePRB2010,von_arx_resonant_2020}. The lowest energy manifold is marked by excitations at about 40 meV and 80 meV. The modes at 80 meV are due to composite magnetic excitations, with dominant transverse character with multiple spin modes, which are allowed among the lowest-energy manifold, due to the spin-orbit coupling at the Ru site (see Fig. 1 in the Supplementary Information). This is consistent with the analysis done for a smaller cluster size ~\cite{von_arx_resonant_2020}. \\
On the other hand, the low-energy peaks at about 40 meV arise from transverse spin modes, i.e. resulting from the $S^{\perp}$ correlation function (symbol $\perp$ refers to the component in the plane orthogonal to the magnetic easy axis), with a mark of having a non-trivial phase factor for the spin modes at nearest neighbor Ru sites (see Supplementary Information). Such non-local (inter-site) spin excitations can be accessible at both the planar and apical oxygens due to the rotation of the octahedra that provides a sublattice structure for the Ru network.  
\\

After having discussed the spin character of the excitations at 40 meV and 80 meV, we consider in more details the orbital nature of these modes. Due to the fact that the energy scales of the spin-orbit coupling and the crystal field are comparable, the lowest-energy excitations are marked by entangled spin-orbital configurations that involve both the spin $S$ and orbital $L$ angular momentum of the $d$-electrons at the Ru site. 
In this regime of crystalline environment, the orbital and spin excitations are thus expected to be anisotropic.
One can then ask whether such orbital modes can be excited in RIXS at the O $K$-edge in the examined range of energy, and which dependence occurs on the direction of the excited orbital moments. This discussion on the orbital character of the low-energy modes will help to account for the polarization and incident angle dependence of the RIXS spectra at the planar and apical resonance. 
To this end, in Figs.~\ref{fig:fig4} b, c we report the dynamical orbital response $L(q,\omega)$, resulting from the orbital angular momentum correlations at different time, for the two Ru sites on the examined cluster. 
We consider the possibility of having longitudinal orbital excitations associated with in-plane orbital correlations ($L_{ab}$) and transverse orbital modes resulting from out-of-plane orbital correlations ($L_c$). This separation is related to the anisotropy of the ground state, having orbital moments lying predominantly in the $ab$ plane. Additionally, we analyze both the correlators with equal [opposite] phase at the Ru sites, corresponding to the $q=(0,0)~ [(\pi,0)]$ modes in the dynamical response, respectively. 
As shown in Figs.~\ref{fig:fig4} b, c, we find that the in-plane orbital excitations are mainly accessible at energies around 80 meV. Instead, the out-of-plane orbital modes occur at lower energies, namely in the range of 40 meV and below it. 
\\
Taking into account these results, we can qualitatively discuss the polarization and oxygen resonance dependence of the RIXS spectra. 
The first remark we make regards the energy distribution of the orbital excitations. We find that the peaks at 80 meV have longitudinal character, namely the excitations at those energies can be achieved by inducing a variation of the orbital angular momentum in the $ab$ plane. By contrast, the low-energy structure around 40 meV is mostly due to excitations that are achieved by orbital modes with out-of-plane orbital angular momentum, having transverse character. It is important to point out that these excitations correspond to variation of the orbital angular momentum rather then of the occupation of the $t_{2g}$ orbitals. This is why they occur at lower energy as compared to the crystal-field derived multiplet structure at about 300 meV.
The orbital nature of the excited states below 100 meV can be put in connection with the polarization dependence of the RIXS spectra.
Although not immediate, we observe that the photon electric field polarization aligned in the $ab$ plane ($c$-axis) is mainly related to the variation of the orbital polarization at the Ru site in the corresponding directions. To get a deeper insight about this connection, one has to consider the $d-p$ hybridized configurations of the type $d^4-p^2$/$d^5-p^1$ that are relevant for the transitions at the oxygen $K$-edge. In particular, the $d^5-p^1$ states that enter in the absorption process at the oxygen $K$-edges can filter configurations with different projections of the orbital angular momentum, depending on the occupation of the orbitals in the $t_{2g}$ manifold. For instance, if the single occupied configuration lies in the $xy$ orbital, then the $L_x$ and $L_y$ components of the orbital angular momentum will be more relevant. 
The behavior of the dynamical orbital response can thus be used to account for the observation of RIXS spectra in LV polarization at the apical oxygen, that exhibits a dominant spectral weight at high energies around 80 meV. This is a direct consequence of the fact that the $p_y$ orbitals of apical oxygens are mainly accessible in LV polarization, and that the corresponding hybridized states are related to variation of the in-plane component of the orbital moments at the Ru site. 
On the other hand, for the planar oxygen, the orbitals involved in the hybridization processes can be linked to both in-plane and out-of-plane variation of the orbital polarization at the Ru site. This implies that the RIXS spectra have peaks both around 40 and 80 meV. We speculate that the transfer of spectral weight from the 80 meV to the 40 meV feature, that is obtained at the planar resonance by increasing the transferred momentum, can be related to the variation of the photon polarization direction. At higher momentum, the photon polarization aligns more in the out-of-plane direction and thus, according to the previous arguments, there is a transfer of spectral weight to low energy. Hence, one may conclude that the in-plane (out-of-plane) variation of the orbital polarization at the Ru site is more relevant for the vertical (horizontal) polarization of the photon electric field, respectively. 

In summary, this work presents an ultra-high-resolution RIXS study of the multiband Mott insulator \Ca, resolving the low-energy excitation spectrum and giving direct insight into spin-orbit coupled excitation modes. The measured excitation spectrum and its theoretical modelling explain how the lowest excitations at 40 and 80 meV are highly non-trivial with a strong spin-orbital character and accessible via RIXS at different light polarizations and absorption resonances through moderate spin-orbit coupling. Thus, we can now track the detailed magnetic excitation spectrum and at the same time infer information about the orbital character. This adds crucial information to the magnetic finestructure of this Mott insulator. Additionally, the resolved splitting of the intra-$t_{2g}$ excitation at 350 meV allows for the direct extraction of the spin-orbit coupling. The rich low-energy excitation structure of this compound is also a fingerprint of its delicate interplay between different energy scales, demonstrating the important role of spin-orbit coupling. Therefore, the improved energy resolution in RIXS can offer new insights for a broad range of strongly correlated systems, where different energy scales influence the low-energy electronic structure.

\noindent\textbf{Methods}\\
High quality single crystals of \Ca\ were grown by the floating zone techniques~\cite{FukazawaPhysB00,snakatsujiJSSCHEM2001}. Crystals were aligned ex-situ with x-ray LAUE and cleaved in-situ using a standard top-post technique. Thermal contact to the cryo-manipulator was achieved using the EPO-TEK E4110 silver epoxy cured overnight at 70 \textcelsius~\cite{SutterNatComm2017a}.  Oxygen $K$-edge x-ray absorption and resonant inelastic x-ray scattering (RIXS) were carried out at the I21 beamline at the DIAMOND synchrotron~\cite{Zhou2022}. All data was recorded at a sample temperature of 15 K. Elastic scattering from a carbon tape yields an energy resolution of 12.5~meV half width at half maximum, more than two times better than previous RIXS studies of \Ca~\cite{das_spin-orbital_2018}. As RIXS spectra were collected along the Ru-O bond direction (only), we use tetragonal notation to indicate momentum space despite orthorhombic crystal structure. This choice also facilitates comparison to existing literature on the topic. All RIXS spectra were normalized to the integrated spectral weight of the $dd$ excitations.

\vspace{2mm}
\noindent\textbf{Acknowledgements}\\
K.v.A., and J.C. acknowledge support by the Swiss National Science Foundation through Grant Number 200021-188564.  K.v.A. is grateful for the support from the FAN Research Talent Development Fund - UZH Alumni and thanks the Forschungskredit of the University of Zurich, grant no. [FK-21-105]. Q.W. was supported by the Research Grants Council of Hong Kong (CUHK 24306223), and the CUHK Direct Grant (No. 4053613). Y.S. thanks the Chalmers Area of Advances-Materials Science and the Swedish Research Council (VR) with a starting Grant (Dnr. 2017-05078) for funding. F.F. and R. F. acknowledge support by the Italian Ministry of University and Research (MIUR), under grant PON 2020JZ5N9M. This work was supported in part by the Italian Ministry of Foreign Affairs and International Cooperation, grant number KR23GR06. We acknowledge Diamond Light Source for time on Beamline I21 under proposal MM27638.
 
\vspace{2mm}
\noindent\textbf{Authors contributions}\\
A.V. and R. F. grew the \Ca\ single crystals. K.v.A. Q.W., J.C., M.G-F., S.A., K.Z. and J.C. carried out the RIXS experiments. P.R. and K.v.A analysed the data with support from Q.W. and J.C.. F.F. developed the theoretical model and carried out the calculations with the help of M.C..
K.v.A., F.F, M.C. and J.C. wrote the manuscript with input from all authors.

\vspace{2mm}
\noindent\textbf{Competing interests}\\
The authors declare no competing interests.

\vspace{2mm}
\noindent\textbf{Data and materials availability}\\ 
The data that support the findings of this study are available from the corresponding author upon request.


\begin{thebibliography}{29}%
\makeatletter
\providecommand \@ifxundefined [1]{%
 \@ifx{#1\undefined}
}%
\providecommand \@ifnum [1]{%
 \ifnum #1\expandafter \@firstoftwo
 \else \expandafter \@secondoftwo
 \fi
}%
\providecommand \@ifx [1]{%
 \ifx #1\expandafter \@firstoftwo
 \else \expandafter \@secondoftwo
 \fi
}%
\providecommand \natexlab [1]{#1}%
\providecommand \enquote  [1]{``#1''}%
\providecommand \bibnamefont  [1]{#1}%
\providecommand \bibfnamefont [1]{#1}%
\providecommand \citenamefont [1]{#1}%
\providecommand \href@noop [0]{\@secondoftwo}%
\providecommand \href [0]{\begingroup \@sanitize@url \@href}%
\providecommand \@href[1]{\@@startlink{#1}\@@href}%
\providecommand \@@href[1]{\endgroup#1\@@endlink}%
\providecommand \@sanitize@url [0]{\catcode `\\12\catcode `\$12\catcode `\&12\catcode `\#12\catcode `\^12\catcode `\_12\catcode `\%12\relax}%
\providecommand \@@startlink[1]{}%
\providecommand \@@endlink[0]{}%
\providecommand \url  [0]{\begingroup\@sanitize@url \@url }%
\providecommand \@url [1]{\endgroup\@href {#1}{\urlprefix }}%
\providecommand \urlprefix  [0]{URL }%
\providecommand \Eprint [0]{\href }%
\providecommand \doibase [0]{https://doi.org/}%
\providecommand \selectlanguage [0]{\@gobble}%
\providecommand \bibinfo  [0]{\@secondoftwo}%
\providecommand \bibfield  [0]{\@secondoftwo}%
\providecommand \translation [1]{[#1]}%
\providecommand \BibitemOpen [0]{}%
\providecommand \bibitemStop [0]{}%
\providecommand \bibitemNoStop [0]{.\EOS\space}%
\providecommand \EOS [0]{\spacefactor3000\relax}%
\providecommand \BibitemShut  [1]{\csname bibitem#1\endcsname}%
\let\auto@bib@innerbib\@empty
\bibitem [{\citenamefont {Witczak-Krempa}\ \emph {et~al.}(2014)\citenamefont {Witczak-Krempa}, \citenamefont {Chen}, \citenamefont {Kim},\ and\ \citenamefont {Balents}}]{WitczakKrempa2014}%
  \BibitemOpen
  \bibfield  {author} {\bibinfo {author} {\bibfnamefont {W.}~\bibnamefont {Witczak-Krempa}}, \bibinfo {author} {\bibfnamefont {G.}~\bibnamefont {Chen}}, \bibinfo {author} {\bibfnamefont {Y.~B.}\ \bibnamefont {Kim}},\ and\ \bibinfo {author} {\bibfnamefont {L.}~\bibnamefont {Balents}},\ }\href {https://doi.org/10.1146/annurev-conmatphys-020911-125138} {\bibfield  {journal} {\bibinfo  {journal} {Annual Review of Condensed Matter Physics}\ }\textbf {\bibinfo {volume} {5}},\ \bibinfo {pages} {57} (\bibinfo {year} {2014})}\BibitemShut {NoStop}%
\bibitem [{\citenamefont {Kim}\ \emph {et~al.}(2008)\citenamefont {Kim}, \citenamefont {Jin}, \citenamefont {Moon}, \citenamefont {Kim}, \citenamefont {Park}, \citenamefont {Leem}, \citenamefont {Yu}, \citenamefont {Noh}, \citenamefont {Kim}, \citenamefont {Oh}, \citenamefont {Park}, \citenamefont {Durairaj}, \citenamefont {Cao},\ and\ \citenamefont {Rotenberg}}]{KimPRL2008}%
  \BibitemOpen
  \bibfield  {author} {\bibinfo {author} {\bibfnamefont {B.~J.}\ \bibnamefont {Kim}}, \bibinfo {author} {\bibfnamefont {H.}~\bibnamefont {Jin}}, \bibinfo {author} {\bibfnamefont {S.~J.}\ \bibnamefont {Moon}}, \bibinfo {author} {\bibfnamefont {J.-Y.}\ \bibnamefont {Kim}}, \bibinfo {author} {\bibfnamefont {B.-G.}\ \bibnamefont {Park}}, \bibinfo {author} {\bibfnamefont {C.~S.}\ \bibnamefont {Leem}}, \bibinfo {author} {\bibfnamefont {J.}~\bibnamefont {Yu}}, \bibinfo {author} {\bibfnamefont {T.~W.}\ \bibnamefont {Noh}}, \bibinfo {author} {\bibfnamefont {C.}~\bibnamefont {Kim}}, \bibinfo {author} {\bibfnamefont {S.-J.}\ \bibnamefont {Oh}}, \bibinfo {author} {\bibfnamefont {J.-H.}\ \bibnamefont {Park}}, \bibinfo {author} {\bibfnamefont {V.}~\bibnamefont {Durairaj}}, \bibinfo {author} {\bibfnamefont {G.}~\bibnamefont {Cao}},\ and\ \bibinfo {author} {\bibfnamefont {E.}~\bibnamefont {Rotenberg}},\ }\href {https://doi.org/10.1103/PhysRevLett.101.076402} {\bibfield  {journal} {\bibinfo  {journal} {Phys. Rev. Lett.}\
  }\textbf {\bibinfo {volume} {101}},\ \bibinfo {pages} {076402} (\bibinfo {year} {2008})}\BibitemShut {NoStop}%
\bibitem [{\citenamefont {Pesin}\ and\ \citenamefont {Balents}(2010)}]{Pesin2010}%
  \BibitemOpen
  \bibfield  {author} {\bibinfo {author} {\bibfnamefont {D.}~\bibnamefont {Pesin}}\ and\ \bibinfo {author} {\bibfnamefont {L.}~\bibnamefont {Balents}},\ }\href {https://doi.org/10.1038/nphys1606} {\bibfield  {journal} {\bibinfo  {journal} {Nature Physics}\ }\textbf {\bibinfo {volume} {6}},\ \bibinfo {pages} {376} (\bibinfo {year} {2010})}\BibitemShut {NoStop}%
\bibitem [{\citenamefont {Bernevig}\ \emph {et~al.}(2006)\citenamefont {Bernevig}, \citenamefont {Hughes},\ and\ \citenamefont {Zhang}}]{Bernevig2006}%
  \BibitemOpen
  \bibfield  {author} {\bibinfo {author} {\bibfnamefont {B.~A.}\ \bibnamefont {Bernevig}}, \bibinfo {author} {\bibfnamefont {T.~L.}\ \bibnamefont {Hughes}},\ and\ \bibinfo {author} {\bibfnamefont {S.-C.}\ \bibnamefont {Zhang}},\ }\href {https://doi.org/10.1126/science.1133734} {\bibfield  {journal} {\bibinfo  {journal} {Science}\ }\textbf {\bibinfo {volume} {314}},\ \bibinfo {pages} {1757} (\bibinfo {year} {2006})}\BibitemShut {NoStop}%
\bibitem [{\citenamefont {Chen}\ and\ \citenamefont {Balents}(2008)}]{ChenPRB2008}%
  \BibitemOpen
  \bibfield  {author} {\bibinfo {author} {\bibfnamefont {G.}~\bibnamefont {Chen}}\ and\ \bibinfo {author} {\bibfnamefont {L.}~\bibnamefont {Balents}},\ }\href {https://doi.org/10.1103/PhysRevB.78.094403} {\bibfield  {journal} {\bibinfo  {journal} {Phys. Rev. B}\ }\textbf {\bibinfo {volume} {78}},\ \bibinfo {pages} {094403} (\bibinfo {year} {2008})}\BibitemShut {NoStop}%
\bibitem [{\citenamefont {Kunkem\"oller}\ \emph {et~al.}(2015)\citenamefont {Kunkem\"oller}, \citenamefont {Khomskii}, \citenamefont {Steffens}, \citenamefont {Piovano}, \citenamefont {Nugroho},\ and\ \citenamefont {Braden}}]{KunkemollerPRL2015}%
  \BibitemOpen
  \bibfield  {author} {\bibinfo {author} {\bibfnamefont {S.}~\bibnamefont {Kunkem\"oller}}, \bibinfo {author} {\bibfnamefont {D.}~\bibnamefont {Khomskii}}, \bibinfo {author} {\bibfnamefont {P.}~\bibnamefont {Steffens}}, \bibinfo {author} {\bibfnamefont {A.}~\bibnamefont {Piovano}}, \bibinfo {author} {\bibfnamefont {A.~A.}\ \bibnamefont {Nugroho}},\ and\ \bibinfo {author} {\bibfnamefont {M.}~\bibnamefont {Braden}},\ }\href {https://doi.org/10.1103/PhysRevLett.115.247201} {\bibfield  {journal} {\bibinfo  {journal} {Phys. Rev. Lett.}\ }\textbf {\bibinfo {volume} {115}},\ \bibinfo {pages} {247201} (\bibinfo {year} {2015})}\BibitemShut {NoStop}%
\bibitem [{\citenamefont {Ament}\ \emph {et~al.}(2011)\citenamefont {Ament}, \citenamefont {van Veenendaal}, \citenamefont {Devereaux}, \citenamefont {Hill},\ and\ \citenamefont {van~den Brink}}]{amentRMP2011}%
  \BibitemOpen
  \bibfield  {author} {\bibinfo {author} {\bibfnamefont {L.~J.~P.}\ \bibnamefont {Ament}}, \bibinfo {author} {\bibfnamefont {M.}~\bibnamefont {van Veenendaal}}, \bibinfo {author} {\bibfnamefont {T.~P.}\ \bibnamefont {Devereaux}}, \bibinfo {author} {\bibfnamefont {J.~P.}\ \bibnamefont {Hill}},\ and\ \bibinfo {author} {\bibfnamefont {J.}~\bibnamefont {van~den Brink}},\ }\href {https://doi.org/10.1103/RevModPhys.83.705} {\bibfield  {journal} {\bibinfo  {journal} {Rev. Mod. Phys.}\ }\textbf {\bibinfo {volume} {83}},\ \bibinfo {pages} {705} (\bibinfo {year} {2011})}\BibitemShut {NoStop}%
\bibitem [{\citenamefont {Zhou}\ \emph {et~al.}(2022)\citenamefont {Zhou}, \citenamefont {Walters}, \citenamefont {Garcia-Fernandez}, \citenamefont {Rice}, \citenamefont {Hand}, \citenamefont {Nag}, \citenamefont {Li}, \citenamefont {Agrestini}, \citenamefont {Garland}, \citenamefont {Wang}, \citenamefont {Alcock}, \citenamefont {Nistea}, \citenamefont {Nutter}, \citenamefont {Rubies}, \citenamefont {Knap}, \citenamefont {Gaughran}, \citenamefont {Yuan}, \citenamefont {Chang}, \citenamefont {Emmins},\ and\ \citenamefont {Howell}}]{Zhou2022}%
  \BibitemOpen
  \bibfield  {author} {\bibinfo {author} {\bibfnamefont {K.-J.}\ \bibnamefont {Zhou}}, \bibinfo {author} {\bibfnamefont {A.}~\bibnamefont {Walters}}, \bibinfo {author} {\bibfnamefont {M.}~\bibnamefont {Garcia-Fernandez}}, \bibinfo {author} {\bibfnamefont {T.}~\bibnamefont {Rice}}, \bibinfo {author} {\bibfnamefont {M.}~\bibnamefont {Hand}}, \bibinfo {author} {\bibfnamefont {A.}~\bibnamefont {Nag}}, \bibinfo {author} {\bibfnamefont {J.}~\bibnamefont {Li}}, \bibinfo {author} {\bibfnamefont {S.}~\bibnamefont {Agrestini}}, \bibinfo {author} {\bibfnamefont {P.}~\bibnamefont {Garland}}, \bibinfo {author} {\bibfnamefont {H.}~\bibnamefont {Wang}}, \bibinfo {author} {\bibfnamefont {S.}~\bibnamefont {Alcock}}, \bibinfo {author} {\bibfnamefont {I.}~\bibnamefont {Nistea}}, \bibinfo {author} {\bibfnamefont {B.}~\bibnamefont {Nutter}}, \bibinfo {author} {\bibfnamefont {N.}~\bibnamefont {Rubies}}, \bibinfo {author} {\bibfnamefont {G.}~\bibnamefont {Knap}}, \bibinfo {author} {\bibfnamefont {M.}~\bibnamefont {Gaughran}},
  \bibinfo {author} {\bibfnamefont {F.}~\bibnamefont {Yuan}}, \bibinfo {author} {\bibfnamefont {P.}~\bibnamefont {Chang}}, \bibinfo {author} {\bibfnamefont {J.}~\bibnamefont {Emmins}},\ and\ \bibinfo {author} {\bibfnamefont {G.}~\bibnamefont {Howell}},\ }\href {https://doi.org/10.1107/s1600577522000601} {\bibfield  {journal} {\bibinfo  {journal} {Journal of Synchrotron Radiation}\ }\textbf {\bibinfo {volume} {29}},\ \bibinfo {pages} {563} (\bibinfo {year} {2022})}\BibitemShut {NoStop}%
\bibitem [{\citenamefont {Gorelov}\ \emph {et~al.}(2010)\citenamefont {Gorelov}, \citenamefont {Karolak}, \citenamefont {Wehling}, \citenamefont {Lechermann}, \citenamefont {Lichtenstein},\ and\ \citenamefont {Pavarini}}]{GorelovPRL2010}%
  \BibitemOpen
  \bibfield  {author} {\bibinfo {author} {\bibfnamefont {E.}~\bibnamefont {Gorelov}}, \bibinfo {author} {\bibfnamefont {M.}~\bibnamefont {Karolak}}, \bibinfo {author} {\bibfnamefont {T.~O.}\ \bibnamefont {Wehling}}, \bibinfo {author} {\bibfnamefont {F.}~\bibnamefont {Lechermann}}, \bibinfo {author} {\bibfnamefont {A.~I.}\ \bibnamefont {Lichtenstein}},\ and\ \bibinfo {author} {\bibfnamefont {E.}~\bibnamefont {Pavarini}},\ }\href {https://doi.org/10.1103/PhysRevLett.104.226401} {\bibfield  {journal} {\bibinfo  {journal} {Phys. Rev. Lett.}\ }\textbf {\bibinfo {volume} {104}},\ \bibinfo {pages} {226401} (\bibinfo {year} {2010})}\BibitemShut {NoStop}%
\bibitem [{\citenamefont {Han}\ and\ \citenamefont {Millis}(2018)}]{han_lattice_2018}%
  \BibitemOpen
  \bibfield  {author} {\bibinfo {author} {\bibfnamefont {Q.}~\bibnamefont {Han}}\ and\ \bibinfo {author} {\bibfnamefont {A.}~\bibnamefont {Millis}},\ }\href {https://doi.org/10.1103/PhysRevLett.121.067601} {\bibfield  {journal} {\bibinfo  {journal} {Phys. Rev. Lett.}\ }\textbf {\bibinfo {volume} {121}},\ \bibinfo {pages} {067601} (\bibinfo {year} {2018})}\BibitemShut {NoStop}%
\bibitem [{\citenamefont {Ricc{\`{o}}}\ \emph {et~al.}(2018)\citenamefont {Ricc{\`{o}}}, \citenamefont {Kim}, \citenamefont {Tamai}, \citenamefont {Walker}, \citenamefont {Bruno}, \citenamefont {Cucchi}, \citenamefont {Cappelli}, \citenamefont {Besnard}, \citenamefont {Kim}, \citenamefont {Dudin}, \citenamefont {Hoesch}, \citenamefont {Gutmann}, \citenamefont {Georges}, \citenamefont {Perry},\ and\ \citenamefont {Baumberger}}]{Ricc2018}%
  \BibitemOpen
  \bibfield  {author} {\bibinfo {author} {\bibfnamefont {S.}~\bibnamefont {Ricc{\`{o}}}}, \bibinfo {author} {\bibfnamefont {M.}~\bibnamefont {Kim}}, \bibinfo {author} {\bibfnamefont {A.}~\bibnamefont {Tamai}}, \bibinfo {author} {\bibfnamefont {S.~M.}\ \bibnamefont {Walker}}, \bibinfo {author} {\bibfnamefont {F.~Y.}\ \bibnamefont {Bruno}}, \bibinfo {author} {\bibfnamefont {I.}~\bibnamefont {Cucchi}}, \bibinfo {author} {\bibfnamefont {E.}~\bibnamefont {Cappelli}}, \bibinfo {author} {\bibfnamefont {C.}~\bibnamefont {Besnard}}, \bibinfo {author} {\bibfnamefont {T.~K.}\ \bibnamefont {Kim}}, \bibinfo {author} {\bibfnamefont {P.}~\bibnamefont {Dudin}}, \bibinfo {author} {\bibfnamefont {M.}~\bibnamefont {Hoesch}}, \bibinfo {author} {\bibfnamefont {M.~J.}\ \bibnamefont {Gutmann}}, \bibinfo {author} {\bibfnamefont {A.}~\bibnamefont {Georges}}, \bibinfo {author} {\bibfnamefont {R.~S.}\ \bibnamefont {Perry}},\ and\ \bibinfo {author} {\bibfnamefont {F.}~\bibnamefont {Baumberger}},\ }\bibfield  {journal} {\bibinfo
  {journal} {Nature Communications}\ }\textbf {\bibinfo {volume} {9}},\ \href {https://doi.org/10.1038/s41467-018-06945-0} {10.1038/s41467-018-06945-0} (\bibinfo {year} {2018})\BibitemShut {NoStop}%
\bibitem [{\citenamefont {Zhang}\ \emph {et~al.}(2019)\citenamefont {Zhang}, \citenamefont {McLeod}, \citenamefont {Han}, \citenamefont {Chen}, \citenamefont {Bechtel}, \citenamefont {Yao}, \citenamefont {Gilbert~Corder}, \citenamefont {Ciavatti}, \citenamefont {Tao}, \citenamefont {Aronson}, \citenamefont {Carr}, \citenamefont {Martin}, \citenamefont {Sow}, \citenamefont {Yonezawa}, \citenamefont {Nakamura}, \citenamefont {Terasaki}, \citenamefont {Basov}, \citenamefont {Millis}, \citenamefont {Maeno},\ and\ \citenamefont {Liu}}]{ZhangPRX2019}%
  \BibitemOpen
  \bibfield  {author} {\bibinfo {author} {\bibfnamefont {J.}~\bibnamefont {Zhang}}, \bibinfo {author} {\bibfnamefont {A.~S.}\ \bibnamefont {McLeod}}, \bibinfo {author} {\bibfnamefont {Q.}~\bibnamefont {Han}}, \bibinfo {author} {\bibfnamefont {X.}~\bibnamefont {Chen}}, \bibinfo {author} {\bibfnamefont {H.~A.}\ \bibnamefont {Bechtel}}, \bibinfo {author} {\bibfnamefont {Z.}~\bibnamefont {Yao}}, \bibinfo {author} {\bibfnamefont {S.~N.}\ \bibnamefont {Gilbert~Corder}}, \bibinfo {author} {\bibfnamefont {T.}~\bibnamefont {Ciavatti}}, \bibinfo {author} {\bibfnamefont {T.~H.}\ \bibnamefont {Tao}}, \bibinfo {author} {\bibfnamefont {M.}~\bibnamefont {Aronson}}, \bibinfo {author} {\bibfnamefont {G.~L.}\ \bibnamefont {Carr}}, \bibinfo {author} {\bibfnamefont {M.~C.}\ \bibnamefont {Martin}}, \bibinfo {author} {\bibfnamefont {C.}~\bibnamefont {Sow}}, \bibinfo {author} {\bibfnamefont {S.}~\bibnamefont {Yonezawa}}, \bibinfo {author} {\bibfnamefont {F.}~\bibnamefont {Nakamura}}, \bibinfo {author} {\bibfnamefont
  {I.}~\bibnamefont {Terasaki}}, \bibinfo {author} {\bibfnamefont {D.~N.}\ \bibnamefont {Basov}}, \bibinfo {author} {\bibfnamefont {A.~J.}\ \bibnamefont {Millis}}, \bibinfo {author} {\bibfnamefont {Y.}~\bibnamefont {Maeno}},\ and\ \bibinfo {author} {\bibfnamefont {M.}~\bibnamefont {Liu}},\ }\href {https://doi.org/10.1103/PhysRevX.9.011032} {\bibfield  {journal} {\bibinfo  {journal} {Phys. Rev. X}\ }\textbf {\bibinfo {volume} {9}},\ \bibinfo {pages} {011032} (\bibinfo {year} {2019})}\BibitemShut {NoStop}%
\bibitem [{\citenamefont {Nakatsuji}\ \emph {et~al.}(2004)\citenamefont {Nakatsuji}, \citenamefont {Dobrosavljevi\ifmmode~\acute{c}\else \'{c}\fi{}}, \citenamefont {Tanaskovi\ifmmode~\acute{c}\else \'{c}\fi{}}, \citenamefont {Minakata}, \citenamefont {Fukazawa},\ and\ \citenamefont {Maeno}}]{NakatsujiPRL2004}%
  \BibitemOpen
  \bibfield  {author} {\bibinfo {author} {\bibfnamefont {S.}~\bibnamefont {Nakatsuji}}, \bibinfo {author} {\bibfnamefont {V.}~\bibnamefont {Dobrosavljevi\ifmmode~\acute{c}\else \'{c}\fi{}}}, \bibinfo {author} {\bibfnamefont {D.}~\bibnamefont {Tanaskovi\ifmmode~\acute{c}\else \'{c}\fi{}}}, \bibinfo {author} {\bibfnamefont {M.}~\bibnamefont {Minakata}}, \bibinfo {author} {\bibfnamefont {H.}~\bibnamefont {Fukazawa}},\ and\ \bibinfo {author} {\bibfnamefont {Y.}~\bibnamefont {Maeno}},\ }\href {https://doi.org/10.1103/PhysRevLett.93.146401} {\bibfield  {journal} {\bibinfo  {journal} {Phys. Rev. Lett.}\ }\textbf {\bibinfo {volume} {93}},\ \bibinfo {pages} {146401} (\bibinfo {year} {2004})}\BibitemShut {NoStop}%
\bibitem [{\citenamefont {Das}\ \emph {et~al.}(2018)\citenamefont {Das}, \citenamefont {Forte}, \citenamefont {Fittipaldi}, \citenamefont {Fatuzzo}, \citenamefont {Granata}, \citenamefont {Ivashko}, \citenamefont {Horio}, \citenamefont {Schindler}, \citenamefont {Dantz}, \citenamefont {Tseng}, \citenamefont {McNally}, \citenamefont {R\o{}nnow}, \citenamefont {Wan}, \citenamefont {Christensen}, \citenamefont {Pelliciari}, \citenamefont {Olalde-Velasco}, \citenamefont {Kikugawa}, \citenamefont {Neupert}, \citenamefont {Vecchione}, \citenamefont {Schmitt}, \citenamefont {Cuoco},\ and\ \citenamefont {Chang}}]{das_spin-orbital_2018}%
  \BibitemOpen
  \bibfield  {author} {\bibinfo {author} {\bibfnamefont {L.}~\bibnamefont {Das}}, \bibinfo {author} {\bibfnamefont {F.}~\bibnamefont {Forte}}, \bibinfo {author} {\bibfnamefont {R.}~\bibnamefont {Fittipaldi}}, \bibinfo {author} {\bibfnamefont {C.~G.}\ \bibnamefont {Fatuzzo}}, \bibinfo {author} {\bibfnamefont {V.}~\bibnamefont {Granata}}, \bibinfo {author} {\bibfnamefont {O.}~\bibnamefont {Ivashko}}, \bibinfo {author} {\bibfnamefont {M.}~\bibnamefont {Horio}}, \bibinfo {author} {\bibfnamefont {F.}~\bibnamefont {Schindler}}, \bibinfo {author} {\bibfnamefont {M.}~\bibnamefont {Dantz}}, \bibinfo {author} {\bibfnamefont {Y.}~\bibnamefont {Tseng}}, \bibinfo {author} {\bibfnamefont {D.~E.}\ \bibnamefont {McNally}}, \bibinfo {author} {\bibfnamefont {H.~M.}\ \bibnamefont {R\o{}nnow}}, \bibinfo {author} {\bibfnamefont {W.}~\bibnamefont {Wan}}, \bibinfo {author} {\bibfnamefont {N.~B.}\ \bibnamefont {Christensen}}, \bibinfo {author} {\bibfnamefont {J.}~\bibnamefont {Pelliciari}}, \bibinfo {author} {\bibfnamefont
  {P.}~\bibnamefont {Olalde-Velasco}}, \bibinfo {author} {\bibfnamefont {N.}~\bibnamefont {Kikugawa}}, \bibinfo {author} {\bibfnamefont {T.}~\bibnamefont {Neupert}}, \bibinfo {author} {\bibfnamefont {A.}~\bibnamefont {Vecchione}}, \bibinfo {author} {\bibfnamefont {T.}~\bibnamefont {Schmitt}}, \bibinfo {author} {\bibfnamefont {M.}~\bibnamefont {Cuoco}},\ and\ \bibinfo {author} {\bibfnamefont {J.}~\bibnamefont {Chang}},\ }\href {https://doi.org/10.1103/PhysRevX.8.011048} {\bibfield  {journal} {\bibinfo  {journal} {Phys. Rev. X}\ }\textbf {\bibinfo {volume} {8}},\ \bibinfo {pages} {011048} (\bibinfo {year} {2018})}\BibitemShut {NoStop}%
\bibitem [{\citenamefont {Gretarsson}\ \emph {et~al.}(2019)\citenamefont {Gretarsson}, \citenamefont {Suzuki}, \citenamefont {Kim}, \citenamefont {Ueda}, \citenamefont {Krautloher}, \citenamefont {Kim}, \citenamefont {Yava\ifmmode~\mbox{\c{s}}\else \c{s}\fi{}}, \citenamefont {Khaliullin},\ and\ \citenamefont {Keimer}}]{GretarssonPRB2019}%
  \BibitemOpen
  \bibfield  {author} {\bibinfo {author} {\bibfnamefont {H.}~\bibnamefont {Gretarsson}}, \bibinfo {author} {\bibfnamefont {H.}~\bibnamefont {Suzuki}}, \bibinfo {author} {\bibfnamefont {H.}~\bibnamefont {Kim}}, \bibinfo {author} {\bibfnamefont {K.}~\bibnamefont {Ueda}}, \bibinfo {author} {\bibfnamefont {M.}~\bibnamefont {Krautloher}}, \bibinfo {author} {\bibfnamefont {B.~J.}\ \bibnamefont {Kim}}, \bibinfo {author} {\bibfnamefont {H.}~\bibnamefont {Yava\ifmmode~\mbox{\c{s}}\else \c{s}\fi{}}}, \bibinfo {author} {\bibfnamefont {G.}~\bibnamefont {Khaliullin}},\ and\ \bibinfo {author} {\bibfnamefont {B.}~\bibnamefont {Keimer}},\ }\href {https://doi.org/10.1103/PhysRevB.100.045123} {\bibfield  {journal} {\bibinfo  {journal} {Phys. Rev. B}\ }\textbf {\bibinfo {volume} {100}},\ \bibinfo {pages} {045123} (\bibinfo {year} {2019})}\BibitemShut {NoStop}%
\bibitem [{\citenamefont {Fatuzzo}\ \emph {et~al.}(2015)\citenamefont {Fatuzzo}, \citenamefont {Dantz}, \citenamefont {Fatale}, \citenamefont {Olalde-Velasco}, \citenamefont {Shaik}, \citenamefont {Dalla~Piazza}, \citenamefont {Toth}, \citenamefont {Pelliciari}, \citenamefont {Fittipaldi}, \citenamefont {Vecchione}, \citenamefont {Kikugawa}, \citenamefont {Brooks}, \citenamefont {R\o{}nnow}, \citenamefont {Grioni}, \citenamefont {R\"uegg}, \citenamefont {Schmitt},\ and\ \citenamefont {Chang}}]{FatuzzoPRB2015}%
  \BibitemOpen
  \bibfield  {author} {\bibinfo {author} {\bibfnamefont {C.~G.}\ \bibnamefont {Fatuzzo}}, \bibinfo {author} {\bibfnamefont {M.}~\bibnamefont {Dantz}}, \bibinfo {author} {\bibfnamefont {S.}~\bibnamefont {Fatale}}, \bibinfo {author} {\bibfnamefont {P.}~\bibnamefont {Olalde-Velasco}}, \bibinfo {author} {\bibfnamefont {N.~E.}\ \bibnamefont {Shaik}}, \bibinfo {author} {\bibfnamefont {B.}~\bibnamefont {Dalla~Piazza}}, \bibinfo {author} {\bibfnamefont {S.}~\bibnamefont {Toth}}, \bibinfo {author} {\bibfnamefont {J.}~\bibnamefont {Pelliciari}}, \bibinfo {author} {\bibfnamefont {R.}~\bibnamefont {Fittipaldi}}, \bibinfo {author} {\bibfnamefont {A.}~\bibnamefont {Vecchione}}, \bibinfo {author} {\bibfnamefont {N.}~\bibnamefont {Kikugawa}}, \bibinfo {author} {\bibfnamefont {J.~S.}\ \bibnamefont {Brooks}}, \bibinfo {author} {\bibfnamefont {H.~M.}\ \bibnamefont {R\o{}nnow}}, \bibinfo {author} {\bibfnamefont {M.}~\bibnamefont {Grioni}}, \bibinfo {author} {\bibfnamefont {C.}~\bibnamefont {R\"uegg}}, \bibinfo {author}
  {\bibfnamefont {T.}~\bibnamefont {Schmitt}},\ and\ \bibinfo {author} {\bibfnamefont {J.}~\bibnamefont {Chang}},\ }\href {https://doi.org/10.1103/PhysRevB.91.155104} {\bibfield  {journal} {\bibinfo  {journal} {Phys. Rev. B}\ }\textbf {\bibinfo {volume} {91}},\ \bibinfo {pages} {155104} (\bibinfo {year} {2015})}\BibitemShut {NoStop}%
\bibitem [{\citenamefont {Jain}\ \emph {et~al.}(2017)\citenamefont {Jain}, \citenamefont {Krautloher}, \citenamefont {Porras}, \citenamefont {Ryu}, \citenamefont {Chen}, \citenamefont {Abernathy}, \citenamefont {Park}, \citenamefont {Ivanov}, \citenamefont {Chaloupka}, \citenamefont {Khaliullin}, \citenamefont {Keimer},\ and\ \citenamefont {Kim}}]{jain2017higgs}%
  \BibitemOpen
  \bibfield  {author} {\bibinfo {author} {\bibfnamefont {A.}~\bibnamefont {Jain}}, \bibinfo {author} {\bibfnamefont {M.}~\bibnamefont {Krautloher}}, \bibinfo {author} {\bibfnamefont {J.}~\bibnamefont {Porras}}, \bibinfo {author} {\bibfnamefont {G.~H.}\ \bibnamefont {Ryu}}, \bibinfo {author} {\bibfnamefont {D.~P.}\ \bibnamefont {Chen}}, \bibinfo {author} {\bibfnamefont {D.~L.}\ \bibnamefont {Abernathy}}, \bibinfo {author} {\bibfnamefont {J.~T.}\ \bibnamefont {Park}}, \bibinfo {author} {\bibfnamefont {A.}~\bibnamefont {Ivanov}}, \bibinfo {author} {\bibfnamefont {J.}~\bibnamefont {Chaloupka}}, \bibinfo {author} {\bibfnamefont {G.}~\bibnamefont {Khaliullin}}, \bibinfo {author} {\bibfnamefont {B.}~\bibnamefont {Keimer}},\ and\ \bibinfo {author} {\bibfnamefont {B.~J.}\ \bibnamefont {Kim}},\ }\href {https://doi.org/10.1038/nphys4077} {\bibfield  {journal} {\bibinfo  {journal} {Nat. Phys.}\ }\textbf {\bibinfo {volume} {13}},\ \bibinfo {pages} {633} (\bibinfo {year} {2017})}\BibitemShut {NoStop}%
\bibitem [{\citenamefont {von Arx}\ \emph {et~al.}(2020)\citenamefont {von Arx}, \citenamefont {Forte}, \citenamefont {Horio}, \citenamefont {Granata}, \citenamefont {Wang}, \citenamefont {Das}, \citenamefont {Sassa}, \citenamefont {Fittipaldi}, \citenamefont {Fatuzzo}, \citenamefont {Ivashko}, \citenamefont {Tseng}, \citenamefont {Paris}, \citenamefont {Vecchione}, \citenamefont {Schmitt}, \citenamefont {Cuoco},\ and\ \citenamefont {Chang}}]{von_arx_resonant_2020}%
  \BibitemOpen
  \bibfield  {author} {\bibinfo {author} {\bibfnamefont {K.}~\bibnamefont {von Arx}}, \bibinfo {author} {\bibfnamefont {F.}~\bibnamefont {Forte}}, \bibinfo {author} {\bibfnamefont {M.}~\bibnamefont {Horio}}, \bibinfo {author} {\bibfnamefont {V.}~\bibnamefont {Granata}}, \bibinfo {author} {\bibfnamefont {Q.}~\bibnamefont {Wang}}, \bibinfo {author} {\bibfnamefont {L.}~\bibnamefont {Das}}, \bibinfo {author} {\bibfnamefont {Y.}~\bibnamefont {Sassa}}, \bibinfo {author} {\bibfnamefont {R.}~\bibnamefont {Fittipaldi}}, \bibinfo {author} {\bibfnamefont {C.~G.}\ \bibnamefont {Fatuzzo}}, \bibinfo {author} {\bibfnamefont {O.}~\bibnamefont {Ivashko}}, \bibinfo {author} {\bibfnamefont {Y.}~\bibnamefont {Tseng}}, \bibinfo {author} {\bibfnamefont {E.}~\bibnamefont {Paris}}, \bibinfo {author} {\bibfnamefont {A.}~\bibnamefont {Vecchione}}, \bibinfo {author} {\bibfnamefont {T.}~\bibnamefont {Schmitt}}, \bibinfo {author} {\bibfnamefont {M.}~\bibnamefont {Cuoco}},\ and\ \bibinfo {author} {\bibfnamefont {J.}~\bibnamefont
  {Chang}},\ }\href {https://doi.org/10.1103/PhysRevB.102.235104} {\bibfield  {journal} {\bibinfo  {journal} {Phys. Rev. B}\ }\textbf {\bibinfo {volume} {102}},\ \bibinfo {pages} {235104} (\bibinfo {year} {2020})}\BibitemShut {NoStop}%
\bibitem [{\citenamefont {Lin}\ \emph {et~al.}(2020)\citenamefont {Lin}, \citenamefont {Miao}, \citenamefont {Mazzone}, \citenamefont {Gu}, \citenamefont {Nag}, \citenamefont {Walters}, \citenamefont {Garc\'{\i}a-Fern\'andez}, \citenamefont {Barbour}, \citenamefont {Pelliciari}, \citenamefont {Jarrige}, \citenamefont {Oda}, \citenamefont {Kurosawa}, \citenamefont {Momono}, \citenamefont {Zhou}, \citenamefont {Bisogni}, \citenamefont {Liu},\ and\ \citenamefont {Dean}}]{LinPRL2020}%
  \BibitemOpen
  \bibfield  {author} {\bibinfo {author} {\bibfnamefont {J.~Q.}\ \bibnamefont {Lin}}, \bibinfo {author} {\bibfnamefont {H.}~\bibnamefont {Miao}}, \bibinfo {author} {\bibfnamefont {D.~G.}\ \bibnamefont {Mazzone}}, \bibinfo {author} {\bibfnamefont {G.~D.}\ \bibnamefont {Gu}}, \bibinfo {author} {\bibfnamefont {A.}~\bibnamefont {Nag}}, \bibinfo {author} {\bibfnamefont {A.~C.}\ \bibnamefont {Walters}}, \bibinfo {author} {\bibfnamefont {M.}~\bibnamefont {Garc\'{\i}a-Fern\'andez}}, \bibinfo {author} {\bibfnamefont {A.}~\bibnamefont {Barbour}}, \bibinfo {author} {\bibfnamefont {J.}~\bibnamefont {Pelliciari}}, \bibinfo {author} {\bibfnamefont {I.}~\bibnamefont {Jarrige}}, \bibinfo {author} {\bibfnamefont {M.}~\bibnamefont {Oda}}, \bibinfo {author} {\bibfnamefont {K.}~\bibnamefont {Kurosawa}}, \bibinfo {author} {\bibfnamefont {N.}~\bibnamefont {Momono}}, \bibinfo {author} {\bibfnamefont {K.-J.}\ \bibnamefont {Zhou}}, \bibinfo {author} {\bibfnamefont {V.}~\bibnamefont {Bisogni}}, \bibinfo {author} {\bibfnamefont
  {X.}~\bibnamefont {Liu}},\ and\ \bibinfo {author} {\bibfnamefont {M.~P.~M.}\ \bibnamefont {Dean}},\ }\href {https://doi.org/10.1103/PhysRevLett.124.207005} {\bibfield  {journal} {\bibinfo  {journal} {Phys. Rev. Lett.}\ }\textbf {\bibinfo {volume} {124}},\ \bibinfo {pages} {207005} (\bibinfo {year} {2020})}\BibitemShut {NoStop}%
\bibitem [{\citenamefont {Wang}\ \emph {et~al.}(2021)\citenamefont {Wang}, \citenamefont {von Arx}, \citenamefont {Horio}, \citenamefont {Mukkattukavil}, \citenamefont {K\"{u}spert}, \citenamefont {Sassa}, \citenamefont {Schmitt}, \citenamefont {Nag}, \citenamefont {Pyon}, \citenamefont {Takayama}, \citenamefont {Takagi}, \citenamefont {Garcia-Fernandez}, \citenamefont {Zhou},\ and\ \citenamefont {Chang}}]{WangSCIADV2021}%
  \BibitemOpen
  \bibfield  {author} {\bibinfo {author} {\bibfnamefont {Q.}~\bibnamefont {Wang}}, \bibinfo {author} {\bibfnamefont {K.}~\bibnamefont {von Arx}}, \bibinfo {author} {\bibfnamefont {M.}~\bibnamefont {Horio}}, \bibinfo {author} {\bibfnamefont {D.~J.}\ \bibnamefont {Mukkattukavil}}, \bibinfo {author} {\bibfnamefont {J.}~\bibnamefont {K\"{u}spert}}, \bibinfo {author} {\bibfnamefont {Y.}~\bibnamefont {Sassa}}, \bibinfo {author} {\bibfnamefont {T.}~\bibnamefont {Schmitt}}, \bibinfo {author} {\bibfnamefont {A.}~\bibnamefont {Nag}}, \bibinfo {author} {\bibfnamefont {S.}~\bibnamefont {Pyon}}, \bibinfo {author} {\bibfnamefont {T.}~\bibnamefont {Takayama}}, \bibinfo {author} {\bibfnamefont {H.}~\bibnamefont {Takagi}}, \bibinfo {author} {\bibfnamefont {M.}~\bibnamefont {Garcia-Fernandez}}, \bibinfo {author} {\bibfnamefont {K.-J.}\ \bibnamefont {Zhou}},\ and\ \bibinfo {author} {\bibfnamefont {J.}~\bibnamefont {Chang}},\ }\bibfield  {journal} {\bibinfo  {journal} {Science Advances}\ }\textbf {\bibinfo {volume} {7}},\ \href
  {https://doi.org/10.1126/sciadv.abg7394} {10.1126/sciadv.abg7394} (\bibinfo {year} {2021})\BibitemShut {NoStop}%
\bibitem [{\citenamefont {Geondzhian}\ \emph {et~al.}(2020)\citenamefont {Geondzhian}, \citenamefont {Sambri}, \citenamefont {De~Luca}, \citenamefont {Di~Capua}, \citenamefont {Di~Gennaro}, \citenamefont {Betto}, \citenamefont {Rossi}, \citenamefont {Peng}, \citenamefont {Fumagalli}, \citenamefont {Brookes}, \citenamefont {Braicovich}, \citenamefont {Gilmore}, \citenamefont {Ghiringhelli},\ and\ \citenamefont {Salluzzo}}]{GeondzhianPRL2020}%
  \BibitemOpen
  \bibfield  {author} {\bibinfo {author} {\bibfnamefont {A.}~\bibnamefont {Geondzhian}}, \bibinfo {author} {\bibfnamefont {A.}~\bibnamefont {Sambri}}, \bibinfo {author} {\bibfnamefont {G.~M.}\ \bibnamefont {De~Luca}}, \bibinfo {author} {\bibfnamefont {R.}~\bibnamefont {Di~Capua}}, \bibinfo {author} {\bibfnamefont {E.}~\bibnamefont {Di~Gennaro}}, \bibinfo {author} {\bibfnamefont {D.}~\bibnamefont {Betto}}, \bibinfo {author} {\bibfnamefont {M.}~\bibnamefont {Rossi}}, \bibinfo {author} {\bibfnamefont {Y.~Y.}\ \bibnamefont {Peng}}, \bibinfo {author} {\bibfnamefont {R.}~\bibnamefont {Fumagalli}}, \bibinfo {author} {\bibfnamefont {N.~B.}\ \bibnamefont {Brookes}}, \bibinfo {author} {\bibfnamefont {L.}~\bibnamefont {Braicovich}}, \bibinfo {author} {\bibfnamefont {K.}~\bibnamefont {Gilmore}}, \bibinfo {author} {\bibfnamefont {G.}~\bibnamefont {Ghiringhelli}},\ and\ \bibinfo {author} {\bibfnamefont {M.}~\bibnamefont {Salluzzo}},\ }\href {https://doi.org/10.1103/PhysRevLett.125.126401} {\bibfield  {journal} {\bibinfo
  {journal} {Phys. Rev. Lett.}\ }\textbf {\bibinfo {volume} {125}},\ \bibinfo {pages} {126401} (\bibinfo {year} {2020})}\BibitemShut {NoStop}%
\bibitem [{\citenamefont {Souliou}\ \emph {et~al.}(2017)\citenamefont {Souliou}, \citenamefont {Chaloupka}, \citenamefont {Khaliullin}, \citenamefont {Ryu}, \citenamefont {Jain}, \citenamefont {Kim}, \citenamefont {Le~Tacon},\ and\ \citenamefont {Keimer}}]{souliou_raman_2017}%
  \BibitemOpen
  \bibfield  {author} {\bibinfo {author} {\bibfnamefont {S.-M.}\ \bibnamefont {Souliou}}, \bibinfo {author} {\bibfnamefont {J.}~\bibnamefont {Chaloupka}}, \bibinfo {author} {\bibfnamefont {G.}~\bibnamefont {Khaliullin}}, \bibinfo {author} {\bibfnamefont {G.}~\bibnamefont {Ryu}}, \bibinfo {author} {\bibfnamefont {A.}~\bibnamefont {Jain}}, \bibinfo {author} {\bibfnamefont {B.~J.}\ \bibnamefont {Kim}}, \bibinfo {author} {\bibfnamefont {M.}~\bibnamefont {Le~Tacon}},\ and\ \bibinfo {author} {\bibfnamefont {B.}~\bibnamefont {Keimer}},\ }\href {https://doi.org/10.1103/PhysRevLett.119.067201} {\bibfield  {journal} {\bibinfo  {journal} {Phys. Rev. Lett.}\ }\textbf {\bibinfo {volume} {119}},\ \bibinfo {pages} {067201} (\bibinfo {year} {2017})}\BibitemShut {NoStop}%
\bibitem [{\citenamefont {Vale}\ \emph {et~al.}(2019)\citenamefont {Vale}, \citenamefont {Dashwood}, \citenamefont {Paris}, \citenamefont {Veiga}, \citenamefont {Garcia-Fernandez}, \citenamefont {Nag}, \citenamefont {Walters}, \citenamefont {Zhou}, \citenamefont {Pietsch}, \citenamefont {Jesche}, \citenamefont {Gegenwart}, \citenamefont {Coldea}, \citenamefont {Schmitt},\ and\ \citenamefont {McMorrow}}]{ValePRB2019}%
  \BibitemOpen
  \bibfield  {author} {\bibinfo {author} {\bibfnamefont {J.~G.}\ \bibnamefont {Vale}}, \bibinfo {author} {\bibfnamefont {C.~D.}\ \bibnamefont {Dashwood}}, \bibinfo {author} {\bibfnamefont {E.}~\bibnamefont {Paris}}, \bibinfo {author} {\bibfnamefont {L.~S.~I.}\ \bibnamefont {Veiga}}, \bibinfo {author} {\bibfnamefont {M.}~\bibnamefont {Garcia-Fernandez}}, \bibinfo {author} {\bibfnamefont {A.}~\bibnamefont {Nag}}, \bibinfo {author} {\bibfnamefont {A.}~\bibnamefont {Walters}}, \bibinfo {author} {\bibfnamefont {K.-J.}\ \bibnamefont {Zhou}}, \bibinfo {author} {\bibfnamefont {I.-M.}\ \bibnamefont {Pietsch}}, \bibinfo {author} {\bibfnamefont {A.}~\bibnamefont {Jesche}}, \bibinfo {author} {\bibfnamefont {P.}~\bibnamefont {Gegenwart}}, \bibinfo {author} {\bibfnamefont {R.}~\bibnamefont {Coldea}}, \bibinfo {author} {\bibfnamefont {T.}~\bibnamefont {Schmitt}},\ and\ \bibinfo {author} {\bibfnamefont {D.~F.}\ \bibnamefont {McMorrow}},\ }\href {https://doi.org/10.1103/PhysRevB.100.224303} {\bibfield  {journal} {\bibinfo
  {journal} {Phys. Rev. B}\ }\textbf {\bibinfo {volume} {100}},\ \bibinfo {pages} {224303} (\bibinfo {year} {2019})}\BibitemShut {NoStop}%
\bibitem [{\citenamefont {Dashwood}\ \emph {et~al.}(2021)\citenamefont {Dashwood}, \citenamefont {Geondzhian}, \citenamefont {Vale}, \citenamefont {Pakpour-Tabrizi}, \citenamefont {Howard}, \citenamefont {Faure}, \citenamefont {Veiga}, \citenamefont {Meyers}, \citenamefont {Chiuzb\ifmmode~\u{a}\else \u{a}\fi{}ian}, \citenamefont {Nicolaou}, \citenamefont {Jaouen}, \citenamefont {Jackman}, \citenamefont {Nag}, \citenamefont {Garc\'{\i}a-Fern\'andez}, \citenamefont {Zhou}, \citenamefont {Walters}, \citenamefont {Gilmore}, \citenamefont {McMorrow},\ and\ \citenamefont {Dean}}]{Dashwood2021}%
  \BibitemOpen
  \bibfield  {author} {\bibinfo {author} {\bibfnamefont {C.~D.}\ \bibnamefont {Dashwood}}, \bibinfo {author} {\bibfnamefont {A.}~\bibnamefont {Geondzhian}}, \bibinfo {author} {\bibfnamefont {J.~G.}\ \bibnamefont {Vale}}, \bibinfo {author} {\bibfnamefont {A.~C.}\ \bibnamefont {Pakpour-Tabrizi}}, \bibinfo {author} {\bibfnamefont {C.~A.}\ \bibnamefont {Howard}}, \bibinfo {author} {\bibfnamefont {Q.}~\bibnamefont {Faure}}, \bibinfo {author} {\bibfnamefont {L.~S.~I.}\ \bibnamefont {Veiga}}, \bibinfo {author} {\bibfnamefont {D.}~\bibnamefont {Meyers}}, \bibinfo {author} {\bibfnamefont {S.~G.}\ \bibnamefont {Chiuzb\ifmmode~\u{a}\else \u{a}\fi{}ian}}, \bibinfo {author} {\bibfnamefont {A.}~\bibnamefont {Nicolaou}}, \bibinfo {author} {\bibfnamefont {N.}~\bibnamefont {Jaouen}}, \bibinfo {author} {\bibfnamefont {R.~B.}\ \bibnamefont {Jackman}}, \bibinfo {author} {\bibfnamefont {A.}~\bibnamefont {Nag}}, \bibinfo {author} {\bibfnamefont {M.}~\bibnamefont {Garc\'{\i}a-Fern\'andez}}, \bibinfo {author} {\bibfnamefont {K.-J.}\
  \bibnamefont {Zhou}}, \bibinfo {author} {\bibfnamefont {A.~C.}\ \bibnamefont {Walters}}, \bibinfo {author} {\bibfnamefont {K.}~\bibnamefont {Gilmore}}, \bibinfo {author} {\bibfnamefont {D.~F.}\ \bibnamefont {McMorrow}},\ and\ \bibinfo {author} {\bibfnamefont {M.~P.~M.}\ \bibnamefont {Dean}},\ }\href {https://doi.org/10.1103/PhysRevX.11.041052} {\bibfield  {journal} {\bibinfo  {journal} {Phys. Rev. X}\ }\textbf {\bibinfo {volume} {11}},\ \bibinfo {pages} {041052} (\bibinfo {year} {2021})}\BibitemShut {NoStop}%
\bibitem [{\citenamefont {Porter}\ \emph {et~al.}(2018)\citenamefont {Porter}, \citenamefont {Granata}, \citenamefont {Forte}, \citenamefont {Di~Matteo}, \citenamefont {Cuoco}, \citenamefont {Fittipaldi}, \citenamefont {Vecchione},\ and\ \citenamefont {Bombardi}}]{porterPRB2018}%
  \BibitemOpen
  \bibfield  {author} {\bibinfo {author} {\bibfnamefont {D.~G.}\ \bibnamefont {Porter}}, \bibinfo {author} {\bibfnamefont {V.}~\bibnamefont {Granata}}, \bibinfo {author} {\bibfnamefont {F.}~\bibnamefont {Forte}}, \bibinfo {author} {\bibfnamefont {S.}~\bibnamefont {Di~Matteo}}, \bibinfo {author} {\bibfnamefont {M.}~\bibnamefont {Cuoco}}, \bibinfo {author} {\bibfnamefont {R.}~\bibnamefont {Fittipaldi}}, \bibinfo {author} {\bibfnamefont {A.}~\bibnamefont {Vecchione}},\ and\ \bibinfo {author} {\bibfnamefont {A.}~\bibnamefont {Bombardi}},\ }\href {https://doi.org/https://doi.org/10.1103/PhysRevB.98.125142} {\bibfield  {journal} {\bibinfo  {journal} {Phys. Rev. B}\ }\textbf {\bibinfo {volume} {98}},\ \bibinfo {pages} {125142} (\bibinfo {year} {2018})}\BibitemShut {NoStop}%
\bibitem [{\citenamefont {Forte}\ \emph {et~al.}(2010)\citenamefont {Forte}, \citenamefont {Cuoco},\ and\ \citenamefont {Noce}}]{fortePRB2010}%
  \BibitemOpen
  \bibfield  {author} {\bibinfo {author} {\bibfnamefont {F.}~\bibnamefont {Forte}}, \bibinfo {author} {\bibfnamefont {M.}~\bibnamefont {Cuoco}},\ and\ \bibinfo {author} {\bibfnamefont {C.}~\bibnamefont {Noce}},\ }\href {https://doi.org/https://doi.org/10.1103/PhysRevB.82.155104} {\bibfield  {journal} {\bibinfo  {journal} {Phys. Rev. B}\ }\textbf {\bibinfo {volume} {82}},\ \bibinfo {pages} {155104} (\bibinfo {year} {2010})}\BibitemShut {NoStop}%
\bibitem [{\citenamefont {Fukazawa}\ \emph {et~al.}(2000)\citenamefont {Fukazawa}, \citenamefont {Nakatsuji},\ and\ \citenamefont {Maeno}}]{FukazawaPhysB00}%
  \BibitemOpen
  \bibfield  {author} {\bibinfo {author} {\bibfnamefont {H.}~\bibnamefont {Fukazawa}}, \bibinfo {author} {\bibfnamefont {S.}~\bibnamefont {Nakatsuji}},\ and\ \bibinfo {author} {\bibfnamefont {Y.}~\bibnamefont {Maeno}},\ }\href {https://doi.org/https://doi.org/10.1016/S0921-4526(99)00989-8} {\bibfield  {journal} {\bibinfo  {journal} {Physica B}\ }\textbf {\bibinfo {volume} {281}},\ \bibinfo {pages} {613} (\bibinfo {year} {2000})}\BibitemShut {NoStop}%
\bibitem [{\citenamefont {Nakatsuji}\ and\ \citenamefont {Maeno}(2001)}]{snakatsujiJSSCHEM2001}%
  \BibitemOpen
  \bibfield  {author} {\bibinfo {author} {\bibfnamefont {S.}~\bibnamefont {Nakatsuji}}\ and\ \bibinfo {author} {\bibfnamefont {Y.}~\bibnamefont {Maeno}},\ }\href {https://doi.org/10.1006/jssc.2000.8953} {\bibfield  {journal} {\bibinfo  {journal} {Journal of Solid State Chemistry}\ }\textbf {\bibinfo {volume} {156}},\ \bibinfo {pages} {26 } (\bibinfo {year} {2001})}\BibitemShut {NoStop}%
\bibitem [{\citenamefont {Sutter}\ \emph {et~al.}(2017)\citenamefont {Sutter}, \citenamefont {Fatuzzo}, \citenamefont {Moser}, \citenamefont {Kim}, \citenamefont {Fittipaldi}, \citenamefont {Vecchione}, \citenamefont {Granata}, \citenamefont {Sassa}, \citenamefont {Cossalter}, \citenamefont {Gatti}, \citenamefont {Grioni}, \citenamefont {R\o{}nnow}, \citenamefont {Plumb}, \citenamefont {Matt}, \citenamefont {Shi}, \citenamefont {Hoesch}, \citenamefont {Kim}, \citenamefont {Chang}, \citenamefont {Jeng}, \citenamefont {Jozwiak}, \citenamefont {Bostwick}, \citenamefont {Rotenberg}, \citenamefont {Georges}, \citenamefont {Neupert},\ and\ \citenamefont {Chang}}]{SutterNatComm2017a}%
  \BibitemOpen
  \bibfield  {author} {\bibinfo {author} {\bibfnamefont {D.}~\bibnamefont {Sutter}}, \bibinfo {author} {\bibfnamefont {C.}~\bibnamefont {Fatuzzo}}, \bibinfo {author} {\bibfnamefont {S.}~\bibnamefont {Moser}}, \bibinfo {author} {\bibfnamefont {M.}~\bibnamefont {Kim}}, \bibinfo {author} {\bibfnamefont {R.}~\bibnamefont {Fittipaldi}}, \bibinfo {author} {\bibfnamefont {A.}~\bibnamefont {Vecchione}}, \bibinfo {author} {\bibfnamefont {V.}~\bibnamefont {Granata}}, \bibinfo {author} {\bibfnamefont {Y.}~\bibnamefont {Sassa}}, \bibinfo {author} {\bibfnamefont {F.}~\bibnamefont {Cossalter}}, \bibinfo {author} {\bibfnamefont {G.}~\bibnamefont {Gatti}}, \bibinfo {author} {\bibfnamefont {M.}~\bibnamefont {Grioni}}, \bibinfo {author} {\bibfnamefont {H.~M.}\ \bibnamefont {R\o{}nnow}}, \bibinfo {author} {\bibfnamefont {N.~C.}\ \bibnamefont {Plumb}}, \bibinfo {author} {\bibfnamefont {C.~E.}\ \bibnamefont {Matt}}, \bibinfo {author} {\bibfnamefont {M.}~\bibnamefont {Shi}}, \bibinfo {author} {\bibfnamefont {M.}~\bibnamefont
  {Hoesch}}, \bibinfo {author} {\bibfnamefont {T.~K.}\ \bibnamefont {Kim}}, \bibinfo {author} {\bibfnamefont {T.~R.}\ \bibnamefont {Chang}}, \bibinfo {author} {\bibfnamefont {H.~T.}\ \bibnamefont {Jeng}}, \bibinfo {author} {\bibfnamefont {C.}~\bibnamefont {Jozwiak}}, \bibinfo {author} {\bibfnamefont {A.}~\bibnamefont {Bostwick}}, \bibinfo {author} {\bibfnamefont {E.}~\bibnamefont {Rotenberg}}, \bibinfo {author} {\bibfnamefont {A.}~\bibnamefont {Georges}}, \bibinfo {author} {\bibfnamefont {T.}~\bibnamefont {Neupert}},\ and\ \bibinfo {author} {\bibfnamefont {J.}~\bibnamefont {Chang}},\ }\href {https://doi.org/10.1038/ncomms15176} {\bibfield  {journal} {\bibinfo  {journal} {Nat. Comm.}\ }\textbf {\bibinfo {volume} {8}},\ \bibinfo {pages} {15176} (\bibinfo {year} {2017})}\BibitemShut {NoStop}%
\end{thebibliography}
%

\end{document}